\documentclass[preprintnumbers,amsmath,amssymbm,prd]{revtex4}
\usepackage{epsfig}
\usepackage{graphicx}

\begin{document}
\title{Asymptotic late-time tails of massive spin-2 fields}
\author{Shahar Hod}
\address{The Ruppin Academic Center, Emeq Hefer 40250, Israel}
\address{ }
\address{The Hadassah Institute, Jerusalem 91010, Israel}
\date{\today}

\begin{abstract}
\ \ \ The late-time dynamics of massive spin-2 fields in flat and
curved spacetimes are studied analytically. We find that the time
evolutions of the massive fields are characterized by oscillatory
power-law decaying tails at asymptotically late times. In a flat
spacetime the decaying exponent depends on the multipole number and
the parity of the mode. In the curved Schwarzschild black-hole
spacetime the decaying exponent is found to be universal.
\end{abstract}
%\bigskip
\maketitle

%]

\section{Introduction}

Massive bosonic fields with arbitrary spins are predicted by many
beyond-standard-model theories \cite{Bou,Has,Rham,RoDo,CarPan}. In
addition, several extensions of the general theory of relativity
which include massive mediator fields have been proposed in the
literature, see \cite{Bou,Has,Rham,RoDo,CarPan} and references
therein. It is therefore of physical interest to explore the
characteristic dynamics of these massive fields.

One of the most remarkable features of the dynamics of massive
fields is the development of asymptotic ($t\to\infty$) tails. At
late times, after the passage of the primary pulse, massive wave
fields do not cut off sharply. Instead, they tend to die off gently,
leaving behind asymptotically late-time decaying tails
\cite{Mor,Hodml,KonZed}. Massive fields are unique in that their
{\it flat}-space dynamics are characterized by non-vanishing tails.
Massless fields, on the other hand, are characterized by tail-free
propagation in a flat spacetime. [For the well-studied phenomena of
massless wave tails in curved spacetimes, see \cite{Tails} and
references therein.]

The asymptotic tails which characterize the late-time dynamics of
massive fields are a direct consequence of the fact that different
frequencies which compose the massive wave packet have different
phase velocities. For example, the flat-space dynamics of a
spherically symmetric spin-0 (scalar) field of mass $\mu$ is
governed by the Klein-Gordon equation
\begin{equation}\label{Eq1}
\Big({{\partial^2}\over{\partial t^2}}-{{\partial^2}\over{\partial
r^2}}+\mu^2\Big)\Psi_0(t,r)=0\  .
\end{equation}
[We use units in which $G=c=\hbar=1$. In these units $\mu$ has the
dimensions of $\text{(length)}^{-1}$.] For a plane wave of the form
$\Psi(t,r)\sim e^{i(kr-\omega t)}$, one finds from (\ref{Eq1}) the
dispersion relation $\omega(k)=\sqrt{k^2+\mu^2}$ which yields the
frequency-dependent phase velocity \cite{Mor}
\begin{equation}\label{Eq2}
v_{\text{phase}}=[1-(\mu/\omega)^2]^{-1/2}\  .
\end{equation}
For a spherically symmetric massive scalar field, it was shown in
\cite{Mor} that the interference between the different components of
the massive wave packet [which are characterized by different phase
velocities, see Eq. (\ref{Eq2})] produces a late-time decaying tail
of the form \cite{Mor}
\begin{equation}\label{Eq3}
\Psi_0(t\gg r)\sim t^{-3/2}\cos(\mu t)\  .
%\ \ \ \text{for} \ \ \t\to\infty\ .
\end{equation}
Here $r$ is the location of the
%initial data and the
observer.
%respectively.

The late-time tails of generic (non-spherical) massive scalar fields
were analyzed in \cite{Hodml}. Resolving the field into spherical
harmonics $\Psi(r,t)=\sum_{lm} \psi_{lm}(r)
Y_{lm}(\theta,\phi)e^{-i\omega t}$ \cite{Notegr}, one obtains the
flat-space Schr\"odinger-like wave equation \cite{Hodml,NoteSch}:
\begin{equation}\label{Eq4}
\Big[{{d^2}\over{dr^2}}+\omega^2-\mu^2-{{l(l+1)}\over{r^2}}\Big]\psi_l(r)=0\
,
\end{equation}
where the multipole number $l$ is a non-negative integer. It was
shown in \cite{Hodml} that the asymptotic late-time dynamics of the
scalar field (\ref{Eq4}) is characterized by an oscillatory decaying
tail of the form
\begin{equation}\label{Eq5}
\Psi_l(t\gg r)\sim t^{-(l+{3\over2})}\cos(\mu t)\  .
%\ \ \ \text{for} \ \ \ t\to\infty\ .
\end{equation}

It is worth pointing out that the oscillatory term $\cos(\mu t)$ in
Eqs. (\ref{Eq3}) and (\ref{Eq5}) reflects the fact that the
late-time ($t\to\infty$) dynamics of a massive field is dominated by
only a small fraction of the frequencies which compose the wave
packet, those with
\begin{equation}\label{Eq6}
|\omega|=\mu-O\big(t^{-1}\big)\  ,
\end{equation}
see \cite{Mor,Hodml} for details.

In the present paper we shall analyze the late-time tails of massive
spin-2 fields. The wave equations for these massive fields were
derived most recently in \cite{CarPan}. We shall first consider the
time evolution of the massive fields in a flat spacetime. We shall
then analyze the late-time dynamics in the curved Schwarzschild
black-hole spacetime.

The problem of analyzing the late-time tails of higher-spin ($s>0$)
massive fields is quite challenging. The main technical problem
arises due to the coupling between the different degrees of freedom
which characterize a higher-spin massive field \cite{RoDo,CarPan}
(this coupling is due to the broken gauge invariance and the
additional longitudinal degree of freedom which characterizes the
dynamics of massive fields \cite{RoDo,CarPan,Notels}). As a
consequence, the dynamics of higher-spin massive fields are governed
by systems of {\it coupled} differential equations \cite{Noteex}. We
shall show, however, that the wave equations can be decoupled in the
asymptotic regime $r\gg\mu^{-1}$. This fact will allow us to study
analytically the asymptotic late-time dynamics of the massive
fields.

\section{The asymptotic late-time tails in a flat spacetime}

We shall now analyze the asymptotic late-time tails of massive
spin-2 fields. The various field modes are classified as even-parity
modes or odd-parity modes according to their behavior under a parity
inversion transformation of the form $\theta\to\pi-\theta$ and
$\phi\to\pi+\pi$ \cite{RoDo,CarPan}. [Here $\theta$ and $\phi$ are
the polar and azimuthal angles, respectively.] Even-parity modes are
multiplied by $(-1)^l$ under this inversion transformation, whereas
odd-parity modes are multiplied by $(-1)^{l+1}$. We shall show below
that the dynamics of these two parity modes are characterized by
different late-time asymptotic behaviors.

\subsection{The odd-parity dipole ($l=1$) mode}

The dynamics of the odd-parity dipole ($l=1$) mode \cite{Notemon2}
of massive spin-2 fields is governed by a single Schr\"odinger-like
wave equation of the form \cite{CarPan,Notecar1}
\begin{equation}\label{Eq7}
\Big({{d^2}\over{dr^2}}+\omega^2-\mu^2-{{6}\over{r^2}}\Big)\psi(r)=0\
.
\end{equation}
Note that Eq. (\ref{Eq7}) is of the form (\ref{Eq4}) with
$l_{\text{eff}}=2$ as the effective multipole parameter. Thus,
taking cognizance of Eq. (\ref{Eq5}), one finds that the asymptotic
late-time tail of the odd-parity dipole ($l=1$) mode is given by
\begin{equation}\label{Eq8}
\Psi(t\gg r)\sim t^{-{7/2}}\cos(\mu t)\  .
%\ \ \ \text{for} \ \ \ t\to\infty\ .
\end{equation}

\subsection{Generic ($l\geq2$) odd-parity modes}

The dynamics of generic ($l\geq2$) odd-parity modes of massive
spin-2 fields is governed by a system of two coupled ordinary
differential equations \cite{CarPan,Notecar2}:
\begin{equation}\label{Eq9}
\Big[{{d^2}\over{dr^2}}+\omega^2-\mu^2-{{l(l+1)+4}\over{r^2}}\Big]\psi_1(r)=
{{2[l(l+1)-2]}\over{r^2}}\psi_2(r)\
\end{equation}
and
\begin{equation}\label{Eq10}
\Big[{{d^2}\over{dr^2}}+\omega^2-\mu^2-{{l(l+1)-2}\over{r^2}}\Big]\psi_2(r)=
{{2}\over{r^2}}\psi_1(r)\  .
\end{equation}

Defining \cite{Notesol1}
\begin{equation}\label{Eq11}
\phi_{\pm}\equiv \psi_1 +{{-3\pm (2l+1)}\over{2}}\psi_2\  ,
\end{equation}
one obtains a pair of {\it decoupled} wave equations:
\begin{equation}\label{Eq12}
\Big[{{d^2}\over{dr^2}}+\omega^2-\mu^2-{{(l+1)(l+2)}\over{r^2}}\Big]\phi_+(r)=0\
\end{equation}
and
\begin{equation}\label{Eq13}
\Big[{{d^2}\over{dr^2}}+\omega^2-\mu^2-{{l(l-1)}\over{r^2}}\Big]\phi_-(r)=0\
.
\end{equation}

Note that Eq. (\ref{Eq12}) is of the form (\ref{Eq4}) with
$l_{\text{eff}}=l+1$ as the effective multipole parameter. Thus,
taking cognizance of Eq. (\ref{Eq5}), one finds that the asymptotic
late-time tail associated with the field $\phi_+$ is given by
\begin{equation}\label{Eq14}
\Phi_+(t\gg r)\sim t^{-(l+{5\over2})}\cos(\mu t)\  .
%\ \ \ \text{for} \ \ \ t\to\infty\ .
\end{equation}
Likewise, note that Eq. (\ref{Eq13}) is of the form (\ref{Eq4}) with
$l_{\text{eff}}=l-1$ as the effective multipole parameter. Thus,
taking cognizance of Eq. (\ref{Eq5}), one finds that the asymptotic
late-time tail associated with the field $\phi_-$ is given by
\begin{equation}\label{Eq15}
\Phi_-(t\gg r)\sim t^{-(l+{1\over2})}\cos(\mu t)\  .
%\ \ \ \text{for} \ \ \ t\to\infty\ .
\end{equation}

\subsection{The even-parity monopole ($l=0$) mode}

The dynamics of the even-parity monopole mode of massive spin-2
fields is governed by a single Schr\"odinger-like wave equation of
the form \cite{CarPan,Notecar3}
\begin{equation}\label{Eq16}
\Big({{d^2}\over{dr^2}}+\omega^2-\mu^2-{{6}\over{r^2}}\Big)\psi(r)=0\
.
\end{equation}
Note that Eq. (\ref{Eq16}) is of the form (\ref{Eq4}) with
$l_{\text{eff}}=2$ as the effective multipole parameter
\cite{Note01}. Thus, taking cognizance of Eq. (\ref{Eq5}), one finds
that the asymptotic late-time tail of the even-parity monopole
($l=0$) mode is given by
\begin{equation}\label{Eq17}
\Psi(t\gg r)\sim t^{-{7/2}}\cos(\mu t)\  .
%\ \ \ \text{for} \ \ \ t\to\infty\ .
\end{equation}

\subsection{The even-parity dipole ($l=1$) mode}

The dynamics of the even-parity dipole mode of massive spin-2 fields
is governed by a system of two coupled ordinary differential
equations \cite{CarPan}. These differential equations are rather
cumbersome [see Eqs. (44)-(45) of \cite{CarPan}] but after some
tedious algebra one finds that, in the asymptotic limit $r\gg
\mu^{-1}$, the two equations can be simplified to yield the
following system of coupled differential equations
\cite{Notereg,Notecar4}:
\begin{equation}\label{Eq18}
\Big({{d^2}\over{dr^2}}+\omega^2-\mu^2-{{8}\over{r^2}}\Big)\psi_1(r)=-{4\over{r^2}}\psi_2(r)\
\end{equation}
and
\begin{equation}\label{Eq19}
\Big({{d^2}\over{dr^2}}+\omega^2-\mu^2-{{6}\over{r^2}}\Big)\psi_2(r)=-{{6}\over{r^2}}\psi_1(r)\
.
\end{equation}

Defining \cite{Notesol2}
\begin{equation}\label{Eq20}
\phi_+\equiv \psi_1+\psi_2\ \ \ \text{and} \ \ \ \phi_-\equiv
\psi_1-{2\over 3}\psi_2\  ,
\end{equation}
one obtains a pair of {\it decoupled} wave equations:
\begin{equation}\label{Eq21}
\Big({{d^2}\over{dr^2}}+\omega^2-\mu^2-{{2}\over{r^2}}\Big)\phi_+(r)=0\
\end{equation}
and
\begin{equation}\label{Eq22}
\Big({{d^2}\over{dr^2}}+\omega^2-\mu^2-{{12}\over{r^2}}\Big)\phi_-(r)=0\
.
\end{equation}

Note that Eq. (\ref{Eq21}) is of the form (\ref{Eq4}) with
$l_{\text{eff}}=1$ as the effective multipole parameter. Thus,
taking cognizance of Eq. (\ref{Eq5}), one finds that the asymptotic
late-time tail associated with the field $\phi_+$ is given by
\begin{equation}\label{Eq23}
\Phi_+(t\gg r\gg \mu^{-1})\sim t^{-{5/2}}\cos(\mu t)\ .
\end{equation}
Likewise, note that Eq. (\ref{Eq22}) is of the form (\ref{Eq4}) with
$l_{\text{eff}}=3$ as the effective multipole parameter. Thus,
taking cognizance of Eq. (\ref{Eq5}), one finds that the asymptotic
late-time tail associated with the field $\phi_-$ is given by
\begin{equation}\label{Eq24}
\Phi_-(t\gg r\gg\mu^{-1})\sim t^{-{9/2}}\cos(\mu t)\ .
\end{equation}

\subsection{Generic ($l\geq2$) even-parity modes}

The dynamics of generic ($l\geq2$) even-parity modes of massive
spin-2 fields is governed by a system of three coupled ordinary
differential equations \cite{CarPan}. These differential equations
are rather lengthy [see Eqs. (38)-(40) of \cite{CarPan}] but after
some tedious algebra one finds that, in the asymptotic limit $r\gg
\mu^{-1}$, the three equations can be simplified to yield the
following system of coupled differential equations
\cite{Notereg,Notecar5}:
\begin{equation}\label{Eq25}
\Big[{{d^2}\over{dr^2}}+\omega^2-\mu^2-{{l(l+1)+6}\over{r^2}}\Big]\psi_1(r)=-{{2l(l+1)}\over{r^2}}\psi_2(r)\
,
\end{equation}
\begin{equation}\label{Eq26}
\Big[{{d^2}\over{dr^2}}+\omega^2-\mu^2-{{l(l+1)+4}\over{r^2}}\Big]\psi_2(r)=-{{6}\over{r^2}}\psi_1(r)
+{{2[l(l+1)-2]}\over{r^2}}\psi_3(r)\  ,
\end{equation}
and
\begin{equation}\label{Eq27}
\Big[{{d^2}\over{dr^2}}+\omega^2-\mu^2-{{l(l+1)-2}\over{r^2}}\Big]\psi_3(r)={{2}\over{r^2}}\psi_2(r)\
.
\end{equation}

Defining \cite{Notesol3}
\begin{equation}\label{Eq28}
\phi_-\equiv \psi_1+{2\over
3}(l+1)\psi_2-{{(l+1)[l(l+1)-2]}\over{3(l-1)}}\psi_3\  ,
\end{equation}
\begin{equation}\label{Eq29}
\phi_+\equiv \psi_1-{2\over
3}l\psi_2-{{l[l(l+1)-2]}\over{3(l+2)}}\psi_3\  ,
\end{equation}
and
\begin{equation}\label{Eq30}
\phi_0\equiv \psi_1+\psi_2+[l(l+1)-2]\psi_3\ ,
\end{equation}
one obtains a system of three {\it decoupled} wave equations:
\begin{equation}\label{Eq31}
\Big[{{d^2}\over{dr^2}}+\omega^2-\mu^2-{{(l-1)(l-2)}\over{r^2}}\Big]\phi_-(r)=0\
,
\end{equation}
\begin{equation}\label{Eq32}
\Big[{{d^2}\over{dr^2}}+\omega^2-\mu^2-{{(l+2)(l+3)}\over{r^2}}\Big]\phi_+(r)=0\
,
\end{equation}
and
\begin{equation}\label{Eq33}
\Big[{{d^2}\over{dr^2}}+\omega^2-\mu^2-{{l(l+1)}\over{r^2}}\Big]\phi_0(r)=0\
\end{equation}

Note that Eq. (\ref{Eq31}) is of the form (\ref{Eq4}) with
$l_{\text{eff}}=l-2$ as the effective multipole parameter. Thus,
taking cognizance of Eq. (\ref{Eq5}), one finds that the asymptotic
late-time tail associated with the field $\phi_-$ is given by
\begin{equation}\label{Eq34}
\Phi_-(t\gg r\gg \mu^{-1})\sim t^{-(l-{1\over 2})}\cos(\mu t)\ .
\end{equation}
Likewise, note that Eq. (\ref{Eq32}) is of the form (\ref{Eq4}) with
$l_{\text{eff}}=l+2$ as the effective multipole parameter. Thus,
taking cognizance of Eq. (\ref{Eq5}), one finds that the asymptotic
late-time tail associated with the field $\phi_+$ is given by
\begin{equation}\label{Eq35}
\Phi_+(t\gg r\gg\mu^{-1})\sim t^{-(l+{7\over 2})}\cos(\mu t)\ .
\end{equation}
Finally, note that Eq. (\ref{Eq33}) is of the form (\ref{Eq4}) with
$l_{\text{eff}}=l$ as the effective multipole parameter. Thus,
taking cognizance of Eq. (\ref{Eq5}), one finds that the asymptotic
late-time tail associated with the field $\phi_0$ is given by
\begin{equation}\label{Eq36}
\Phi_0(t\gg r\gg\mu^{-1})\sim t^{-(l+{3\over 2})}\cos(\mu t)\ .
\end{equation}

\section{The asymptotic late-time tail in the curved Schwarzschild spacetime}

The wave equations for the various modes of massive spin-2 fields in
the Schwarzschild black-hole spacetime are rather lengthy, see Eq.
(36) of \cite{CarPan} for the odd-parity dipole ($l=1$) mode, Eqs.
(32)-(35) of \cite{CarPan} for generic ($l\geq 2$) odd-parity modes,
Eq. (30) of \cite{CarPan} for the even-parity monopole ($l=0$) mode,
Eqs. (44)-(45) of \cite{CarPan} for the even-parity dipole ($l=1$)
mode, and Eqs. (38)-(40) of \cite{CarPan} for generic ($l\geq 2$)
even-parity modes. Remarkably, after some algebra one finds that the
wave equations of the various modes are all characterized by the
{\it same} asymptotic behavior:
\begin{equation}\label{Eq37}
\Big[{{d^2}\over{dr^2}}+\omega^2-\mu^2+{{2M\mu^2}\over{r}}+O\Big({{1}\over{r^2}}\Big)\Big]\psi(r)=0\
\ \ \text{for}\ \ \ r\gg\text{max}\{1/M\mu^2,\mu^{-1}\}\  ,
\end{equation}
where $M$ is the mass of the central black hole. It is important to
note that the new term $2M\mu^2/r$, which represents the curvature
of the black-hole spacetime \cite{Notefl}, dominates over the
coupling terms in the asymptotic limit $r\gg M/(M\mu)^2$
\cite{Notecoupl}.

It was shown in \cite{Koy} that the asymptotic late-time dynamics of
a wave field of the form (\ref{Eq37}) is characterized by an
oscillatory decaying tail of the form
\begin{equation}\label{Eq38}
\Psi(t\gg r\gg 1/M\mu^2)\sim t^{-5/6}\cos(\mu t)\  .
\end{equation}
Thus, the late-time evolution of massive fields in the curved
Schwarzschild black-hole spacetime is universal in the sense that
all modes share the same asymptotic behavior (\ref{Eq38})
\cite{Noteins}. It is worth emphasizing that, for $M\mu\ll 1$, the
flat-space tails of Sec. II dominate the dynamics of the massive
fields in the intermediate asymptotic regime $M\ll t\ll M/(M\mu)^2$,
see \cite{Hodml} for details.

\section{Summary}

The characteristic late-time tails of massive spin-2 fields in flat
and curved spacetimes were studied analytically. The problem of
analyzing the dynamics of higher-spin ($s>0$) massive fields is
technically more challenging than the corresponding problem of
analyzing the time evolution of massless fields. The main technical
difficulty stems from the fact that different degrees of freedom
which characterize the dynamics of massive fields are coupled
\cite{RoDo,CarPan}. As a consequence, the dynamics of higher-spin
massive fields are governed by systems of {\it coupled} ordinary
differential equations \cite{Noteex}. We have shown, however, that
the wave equations can be decoupled in the asymptotic regime
$r\gg\mu^{-1}$. This fact allows one to study analytically the
properties of the asymptotic tails which characterize the late-time
dynamics of the massive fields. The various cases studied and the
corresponding asymptotic late-time tails are summarized in Table
\ref{Table1}.

\begin{table}[htbp]
\centering
\begin{tabular}{|c|c|c|}
\hline \text{Mode(s)} & \ \ $l_{\text{eff}}$ \ \ & \ \text{Asymptotic late-time tail(s)}\ \ \\
\hline
\text{Odd-parity} $l=1$ & \ \ $2$ \ \ & $t^{-{7\over 2}}\cos(\mu t)$ \\
\hline
\ \text{Odd-parity} $l\geq2$\ \ & \ \ $l-1$ \ \ & $t^{-(l+{1\over 2})}\cos(\mu t)$ \\
 & \ \ $l+1$ \ \ & $t^{-(l+{5\over 2})}\cos(\mu t)$ \\
\hline
\text{Even-parity} $l=0$ & \ \ $2$ \ \ & $t^{-{7\over 2}}\cos(\mu t)$ \\
\hline
\text{Even-parity} $l=1$ & \ \ $1$ \ \ & $t^{-{5\over 2}}\cos(\mu t)$ \\
 & \ \ $3$ \ \ & $t^{-{9\over 2}}\cos(\mu t)$ \\
\hline
\ \text{Even-parity} $l\geq2$\ \ & \ \ $l-2$ \ \ & $t^{-(l-{1\over 2})}\cos(\mu t)$ \\
 & \ \ $l$ \ \ & $t^{-(l+{3\over 2})}\cos(\mu t)$ \\
 & \ \ $l+2$ \ \ & $t^{-(l+{7\over 2})}\cos(\mu t)$ \\
\hline
\end{tabular}
\caption{Late time tails of massive spin-2 fields in a flat
spacetime. Here $l$ is the multipole number of the mode and
$l_{\text{eff}}$ is the effective multipole parameter which
determines the asymptotic behavior of the scattering potential in
the $r\to\infty $ limit [see Eq. (\ref{Eq4})]. The asymptotic late
time tail in the curved Schwarzschild black-hole spacetime is given
by $\Psi(t\gg r)\sim t^{-5/6}\cos(\mu t)$ for all modes
\cite{Noteins}.} \label{Table1}
\end{table}

\bigskip
\noindent
{\bf ACKNOWLEDGMENTS}
\bigskip

This research is supported by the Carmel Science Foundation. I would
like to thank Richard Brito, Vitor Cardoso, and Paolo Pani for
helpful correspondence. I would also like to thank Yael Oren, Arbel
M. Ongo and Ayelet B. Lata for helpful discussions.

%\newpage


\begin{thebibliography}{99}
\bibitem{Notegr} Throughout the paper we shall use the capital Greek letters
$\Psi(t,r)$ and $\Phi(t,r)$ to denote the time-dependent fields and
the minuscules Greek letters $\psi(r;\omega)$ and $\phi(r;\omega)$
to denote the radial parts of the fields in the frequency domain.
\bibitem{NoteSch} It is worth emphasizing that the wave equations considered in the present work have a
quadratic dependence of the effective potentials on the frequency,
whereas in the standard Schr\"odinger case there is a linear
dependence on the energy.
\bibitem{Notels} It is worth emphasizing that higher spin ($s>0$) massive fields have longitudinal radiative
degrees of freedom which do not exist in the spectra of their
massless counterparts. In particular, massive higher spin fields
have low multipoles propagating modes with $l<s$, for example: the
vector ($s=1$) field monopole ($l=0$) and the graviton ($s=2$) field
monopole ($l=0$) and dipole ($l=1$) modes (we recall that the
radiative modes of massless fields are characterized by the
inequality $l\geq s$).
\bibitem{Noteex} The only exceptions are the odd-parity dipole ($l=1$) mode
and the even-parity monopole ($l=0$) mode. In these two special
cases the dynamics is governed by a single Schr\"odinger-like wave
equation, see Eqs. (\ref{Eq7}) and (\ref{Eq16}).
\bibitem{Notemon2} Note that the monopole ($l=0$) mode does not exist in the odd-parity
sector of spin-2 perturbations \cite{CarPan}.
\bibitem{Notesol1} Defining $\phi\equiv \psi_1+b\cdot\psi_2$, one
obtains from Eqs. (\ref{Eq9})-(\ref{Eq10}) the wave equation:
$({{d^2}\over{dr^2}}+\omega^2-\mu^2)\phi-{{l(l+1)+4+2b}\over{r^2}}\{\psi_1
+ {{(2+b)[l(l+1)-2]}\over{l(l+1)+4+2b}}\psi_2\}=0$. Demanding the
equality ${{(2+b)[l(l+1)-2]}\over{l(l+1)+4+2b}}=b$, one obtains the
quadratic equation $b^2+3b-[l(l+1)-2]=0$ for the coefficient $b$,
whose solutions are given by Eq. (\ref{Eq11}). The decoupled wave
equations are then given by
$[{{d^2}\over{dr^2}}+\omega^2-\mu^2-{{l(l+1)+4+2b}\over{r^2}}]\phi=0$,
see Eqs. (\ref{Eq12})-(\ref{Eq13}).
\bibitem{Notereg} Here we have used the well known fact that the asymptotic late-time ($t\to\infty$)
tail is dominated by frequencies with $|\omega|\to\mu$
\cite{Mor,Hodml} in order to simplify the equations.
\bibitem{Notesol2} Defining $\phi\equiv \psi_1+b\cdot\psi_2$, one obtains from Eqs.
(\ref{Eq18})-(\ref{Eq19}) the wave equation:
$({{d^2}\over{dr^2}}+\omega^2-\mu^2)\phi-{{8-6b}\over{r^2}}(\psi_1 +
{{-4+6b}\over{8-6b}}\psi_2)=0$. Demanding the equality
${{-4+6b}\over{8-6b}}=b$, one obtains the quadratic equation
$3b^2-b-2=0$ for the coefficient $b$, whose solutions are given by
Eq. (\ref{Eq20}). The decoupled wave equations are then given by
$({{d^2}\over{dr^2}}+\omega^2-\mu^2-{{8-6b}\over{r^2}})\phi=0$, see
Eqs. (\ref{Eq21})-(\ref{Eq22}).
\bibitem{Notesol3} Defining $\phi\equiv \psi_1+b\cdot\psi_2+c\cdot \psi_3$, one
obtains from Eqs. (\ref{Eq25})-(\ref{Eq27}) the wave equation:
$({{d^2}\over{dr^2}}+\omega^2-\mu^2)\phi-{{l(l+1)+6-6b}\over{r^2}}\{\psi_1
+
{{-2l(l+1)+b[l(l+1)+4]+2c}\over{l(l+1)+6-6b}}\psi_2+{{2b[l(l+1)-2]+c[l(l+1)-2]}\over{l(l+1)+6-6b}}\psi_3\}=0$.
Demanding the two equalities
${{-2l(l+1)+b[l(l+1)+4]+2c}\over{l(l+1)+6-6b}}=b$ and
${{2b[l(l+1)-2]+c[l(l+1)-2]}\over{l(l+1)+6-6b}}=c$, one obtains the
cubic equation $9b^3-15b^2+[6-4l(l+1)]b+4l(l+1)=0$ for the
coefficient $b$, whose solutions are given by Eqs.
(\ref{Eq28})-(\ref{Eq30}). The decoupled wave equations are then
given by
$[{{d^2}\over{dr^2}}+\omega^2-\mu^2-{{l(l+1)+6-6b}\over{r^2}}]\phi=0$,
see Eqs. (\ref{Eq31})-(\ref{Eq33}).
\bibitem{Notefl} It is worth emphasizing that this term disappears in the case of a
flat spacetime with $M\to 0$, see Sec. II.
\bibitem{Notecoupl} The coupling terms are of order $O(1/r^2)$, see Eqs. (\ref{Eq9})-(\ref{Eq10}),
(\ref{Eq18})-(\ref{Eq19}), and (\ref{Eq25})-(\ref{Eq27}). On the
other hand, the curvature term is of order $O(M\mu^2/r)$, see Eq.
(\ref{Eq37}). Thus, the curvature term dominates over the coupling
terms in the asymptotic limit $r\gg M/(M\mu)^2$.
\bibitem{Noteins} The only exception is the even-parity monopole ($l=0$) mode. As shown in \cite{CarPan},
this mode produces a Gregory-Laflamme-like instability of the
Schwarzschild black hole in the regime $M\mu\lesssim 0.43$.
\end{thebibliography}
\end{document}